\documentclass[aps,prl,showpacs,twocolumn,superscriptaddress]{revtex4}
\usepackage{amsmath}
\usepackage{amssymb}
\usepackage{epsfig}
\usepackage{color}
\usepackage{graphicx,amsmath}

\begin{document}
\title{Comment on "Correlated impurities and intrinsic spin liquid physics
in the kagome material Herbertsmithite" (T. H. Han et al., \prb
{\bf 94}, 060409(R) (2016))}
\author{V. R. Shaginyan}\email{vrshag@thd.pnpi.spb.ru} \affiliation{Petersburg
Nuclear Physics Institute,  Gatchina, 188300,
Russia}\affiliation{Clark Atlanta University, Atlanta, GA 30314,
USA} \author{M. Ya. Amusia}\affiliation{Racah Institute of Physics,
Hebrew University, Jerusalem 91904, Israel}\affiliation{Ioffe
Physical Technical Institute, RAS, St. Petersburg 194021, Russia}
\author{J.~W.~Clark}
\affiliation{McDonnell Center for the Space Sciences \& Department
of Physics, Washington University, St.~Louis, MO 63130, USA}
\affiliation{Centro de Ci\^encias Matem\'aticas, Universidade de Madeira,       9000-390 Funchal, Madeira, Portugal}
\author{G. S. Japaridze}\affiliation{Clark Atlanta
University, Atlanta, GA 30314, USA}\author{A. Z.
Msezane}\affiliation{Clark Atlanta University, Atlanta, GA
30314, USA}\author{K. G. Popov}\affiliation{Komi Science Center,
Ural Division, RAS, Syktyvkar, 167982, Russia}

\begin{abstract}

Recently Han et al.~\cite{Han} have provided an analysis of the
observed behavior of $\rm ZnCu_{3}(OH)_6Cl_2$ Herbertsmithite
based on a separation of the contributions to its thermodynamic
properties due to impurities from those due to the kagome
lattice. The authors developed an impurity model to account for
the experimental data and claimed that it is compatible with the
presence of a small spin gap in the kagome layers. We argue that
the model they advocate is problematic, conflicting with the
intrinsic properties of $\rm ZnCu_{3}(OH)_6Cl_2$ as observed and
explained in recent experimental and theoretical investigations.
We show that the existence of the gap in the kagome layers is
not in itself of a vital importance, for it does not govern the
thermodynamic and transport properties of $\rm
ZnCu_3(OH)_6Cl_2$. Measurements of heat transport in magnetic
fields could clarify the quantum-critical features of
spin-liquid physics of $\rm ZnCu_{3}(OH)_6Cl_2$.
\end{abstract}


\maketitle

In a frustrated magnet, spins are prevented from forming an
ordered alignment, so they collapse into a liquid-like state
named quantum spin liquid (QSL) even at the temperatures close
to absolute zero. A challenge is to prepare quantum spin liquid
materials in the laboratory and explain their properties. Han et
al. have recently reported results from high-resolution
low-energy inelastic neutron scattering on single-crystal $\rm
ZnCu_{3}(OH)_6Cl_2$ Herbertsmithite with the prospect of
disentangling the effects on the observed properties of this
material due to $\rm Cu$ impurity spins from the effects of the
kagome lattice itself \cite{Han}. Citing single-crystal NMR and
resonant X-ray diffraction measurements indicating that the
impurities are 15\% $\rm Cu$ on triangular $\rm Zn$ intercites
and the kagome planes are fully occupied with $\rm Cu$, the
authors assume that the corresponding impurity system may be
represented as a simple cubic lattice in the dilute limit below
the percolation threshold. They claim that this impurity model
can describe the neutron-scattering measurements and specific
heat data, and they suggest that it is compatible with the
existence of a small spin-gap in the kagome layers
\cite{Han,sc_han,norman}. Then, the model assumes that spin gap
survives under the application of magnetic fields up to 9 T
\cite{sc_han}, while the bulk spin susceptibility $\chi$
exhibits a divergent Curie-like tail, indicating that some of
the $\rm Cu$ spins act like weakly coupled impurities
\cite{Han,sc_han,norman}. According to the impurity model, the
divergent response seen in the dynamic spin susceptibility below
1 meV is due to the impurities \cite{Han,sc_han}.

In this comment we show that 1) The proposed impurity model is
artificial because it is inconsistent with the intrinsic
properties of $\rm ZnCu_{3}(OH)_6Cl_2$ as observed and described
in recent experimental and theoretical studies of the behavior
of its thermodynamic, dynamic and relaxation properties; 2)
Explanation of these properties lies in the physics of the
strongly correlated quantum spin liquid (SCQSL) present in this
system, for the behavior of $\rm ZnCu_{3}(OH)_6Cl_2$ is in fact
similar to that of heavy-fermion metals, with one main exception
--- $\rm ZnCu_{3}(OH)_6Cl_2$ does not support an electrical
current
\cite{pr,shaginyan:2011,shaginyan:2012:A,shaginyan:2011:C,shaginyan:2013:D,comm,hfliq,book,sc2016};
and 3) We demonstrate the proposed impurity model \cite{Han}
cannot describe prior neutron, NMR, and specific heat data
obtained in measurements in magnetic fields; it is also
impossible to isolate the contributions coming from the
impurities and the kagome plains. We conclude by recommending
that measurements of heat transport in magnetic fields $B$ be
carried out; they could be crucial in revealing the mechanisms
involved \cite{shaginyan:2011:C,shaginyan:2013:D,book}.

\begin{figure} [! ht]
\begin{center}
\includegraphics [width=0.38\textwidth]{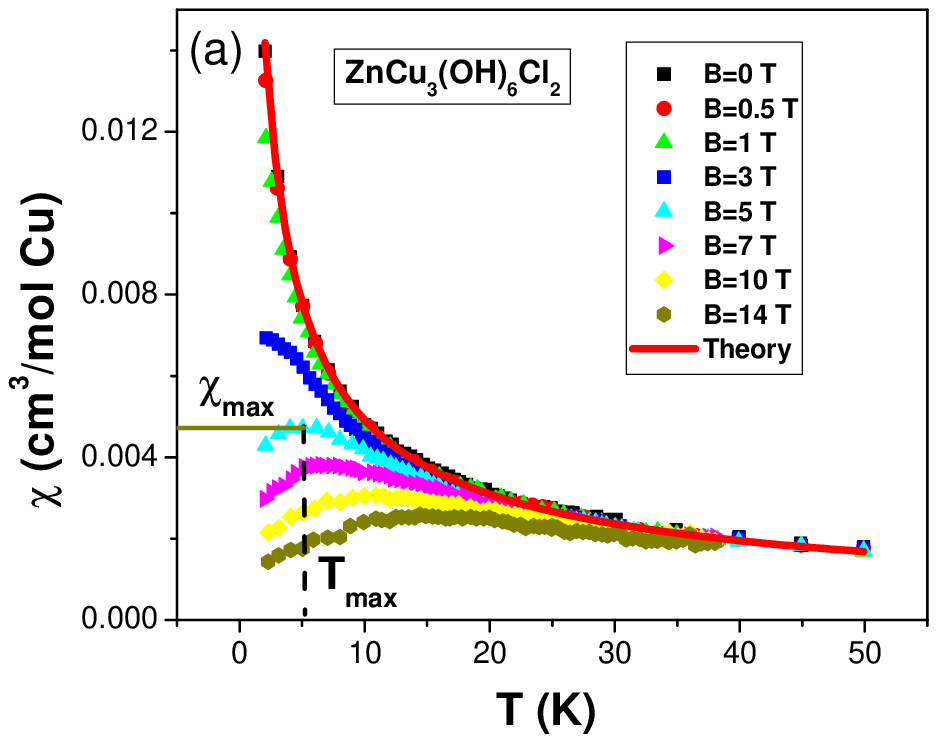}
\includegraphics [width=0.38\textwidth]{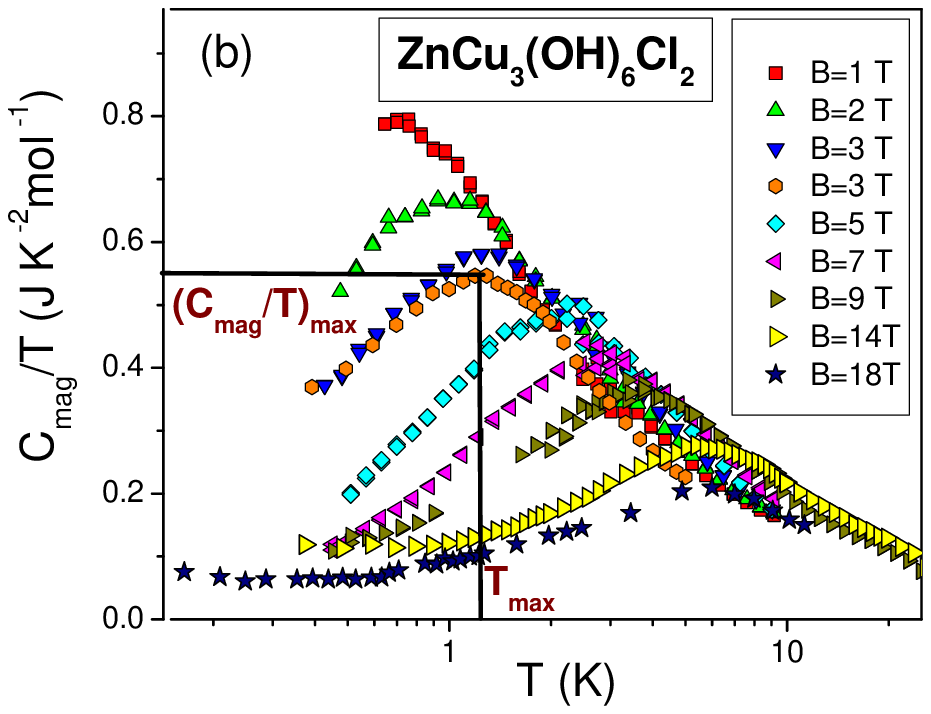}
\includegraphics [width=0.38\textwidth]{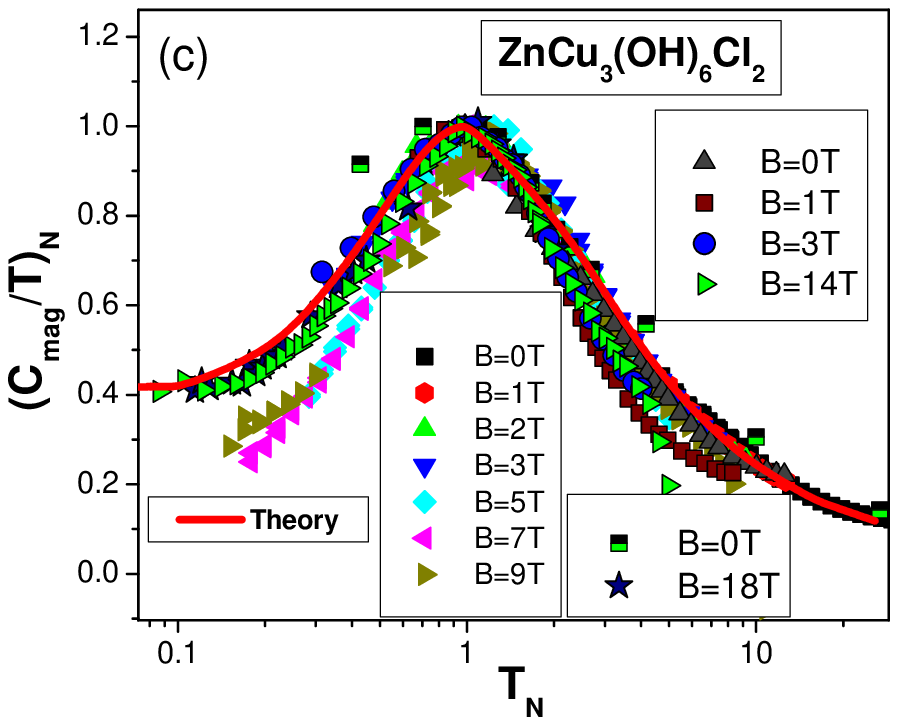}
\end{center}
\vspace*{-0.8cm} \caption{Panel (a): Measured temperature
dependence of the magnetic susceptibility $\chi$ of $\rm
ZnCu_3(OH)_6Cl_2$ from Ref.~\cite{herb3} at magnetic fields
shown in the legend.  Illustrative values of $\chi_{\rm max}$
and $T_{\rm max}$ at $B=3$ T are also shown. A theoretical
prediction at $B=0$ is plotted as the solid curve, which
represents $\chi(T)\propto T^{-\alpha}$ with $\alpha=2/3$
\cite{shaginyan:2011,book}. Panel (b): Specific heat $C_{\rm
mag}/T$ measured on powder \cite{herb2} and single-crystal
\cite{herb2,herb,t_han:2012,t_han:2014} samples of
Herbertsmithite, { is} displayed as a function of temperature
$T$ for fields $B$ shown in the legend. Panel (c): Normalized
specific heat $(C_{\rm mag}/T)_N$ versus normalized temperature
$T_N$ as a function of { $B$} field values shown in the legend
\cite{hfliq,book}. The theoretical result from
Ref.~\cite{shaginyan:2011,book}, represented by the solid curve,
traces the scaling behavior of the effective mass.} \label{fig1}
\end{figure}
\begin{figure} [! ht]
\begin{center}
\includegraphics [width=0.37\textwidth]{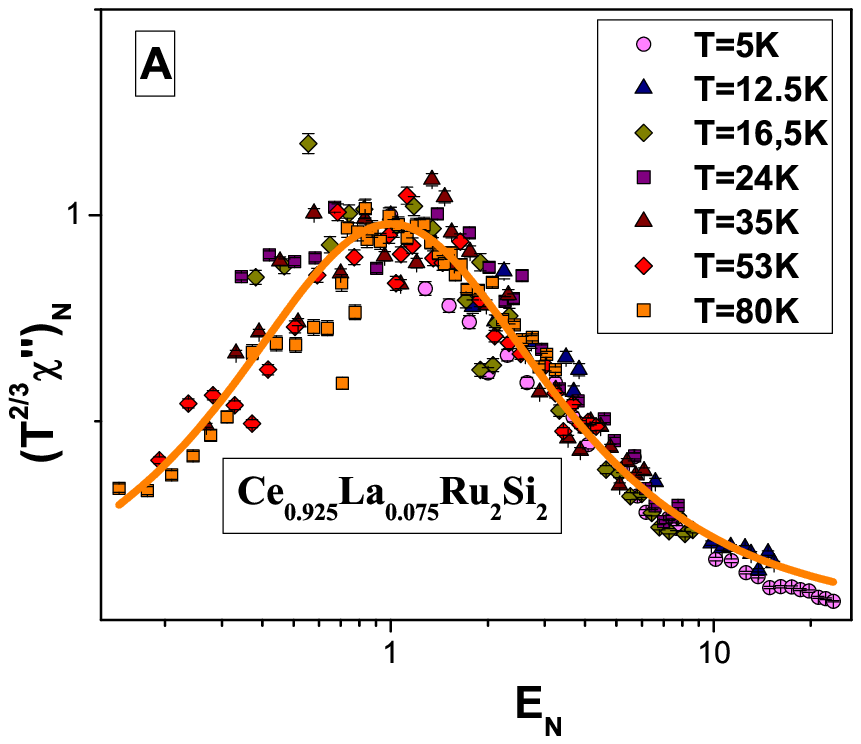}
\includegraphics [width=0.37\textwidth]{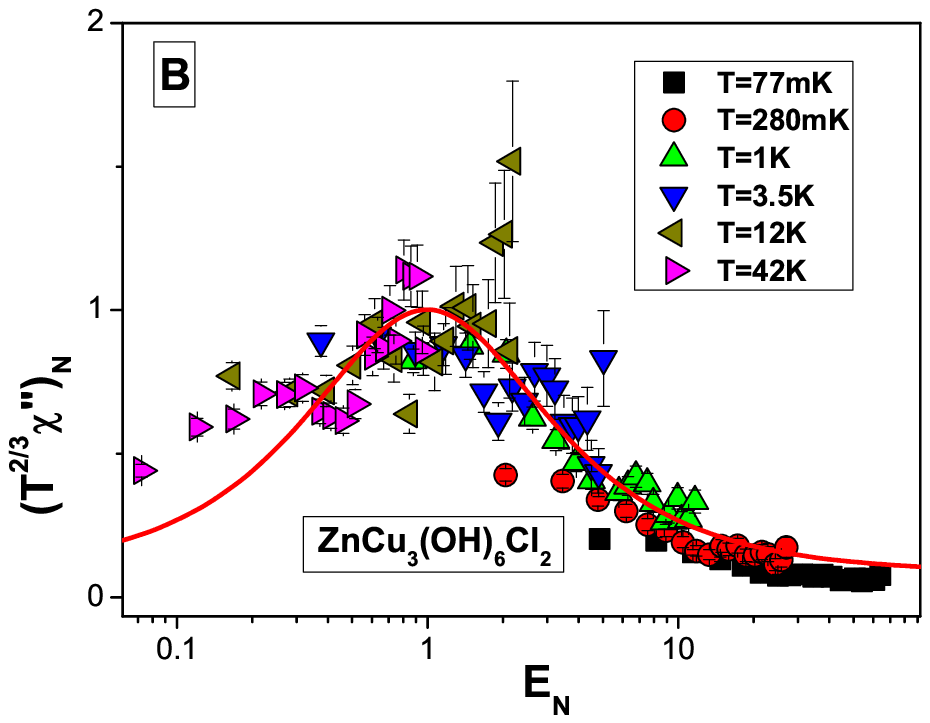}
\includegraphics [width=0.37\textwidth]{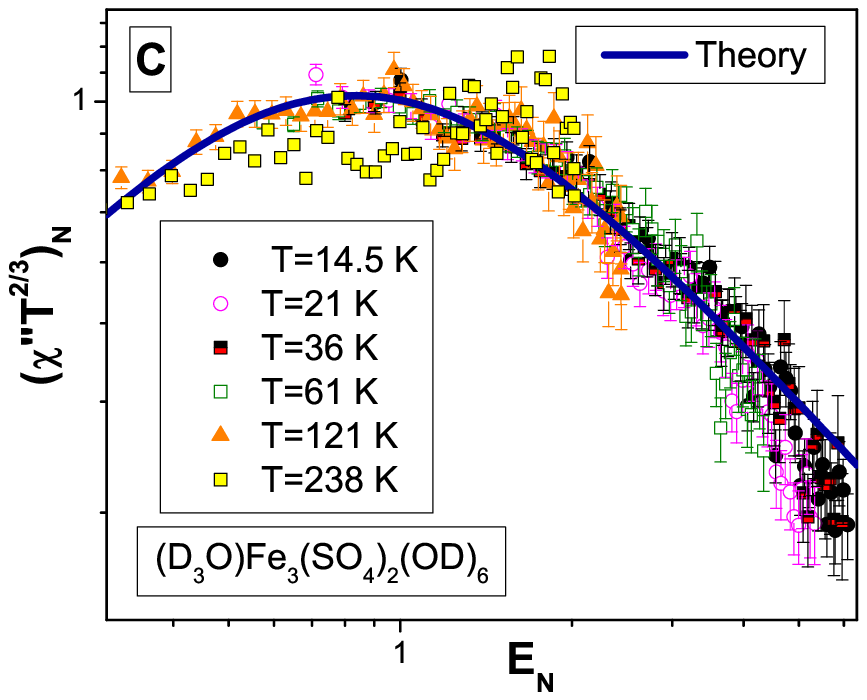}
\end{center}
\caption{Scaling behavior of the normalized dynamic spin
susceptibility $(T^{2/3}\chi'')_N$ for three materials. Panel
{\bf A}: $(T^{2/3}\chi'')_N$ plotted against the dimensionless
variable $E_N$. Data are extracted from measurements on the
heavy-fermion metal $\rm Ce_{0.925}La_{0.075}Ru_2Si_2$
\cite{knafo:2004}. Panel {\bf B}: $(T^{2/3}\chi'')_N$ versus
$E_N$. Data are extracted from measurements on Herbertsmithite
$\rm ZnCu_3(OH)_6Cl_2$ \cite{herb3}. Panel {\bf C}:
$(T^{2/3}\chi'')_N$ versus $E_N$. Data are extracted from
measurements on the deuteronium jarosite $\rm
(D_3O)Fe_3(SO_4)_2(OD)_6$ \cite{faak:2008}. Solid curve:
Theoretical prediction based on Eq.~\eqref{SCHIN}.
\cite{shaginyan:2012:A}.}\label{fig2}
\end{figure}
To examine the model of Han et al. in a broader context, we
first refer to the experimental behavior of the thermodynamic
properties of Herbertsmithite as summarized in Fig.~\ref{fig1}.
It is obvious from Fig.~\ref{fig1}(a) that the magnetic
susceptibility $\chi$ diverges in magnetic fields $B\leq 1$ T
and that the Landau Fermi liquid (LFL) behavior is demonstrated
at least for $B\geq 3$ T and low temperatures $T$; at such
temperatures and magnetic fields the impurities become fully
polarized and hence do not contribute to $\chi$
\cite{herb3,shaginyan:2011:C,book}. Corresponding behavior
follows from Fig.~\ref{fig1}(b); it is seen that LFL behavior of
the heat capacity $C_{\rm mag}/T$ emerges under application of
the same fields. Consequently, we conclude that at $B\geq 3$ T
and low $T$ the contributions to both $\chi$ and $C_{\rm mag}/T$
from the impurities are negligible; rather, one could expect
that they are dominated by the kagome lattice exhibiting a spin
gap in the kagome layers \cite{Han,sc_han,norman}. Thus, one
would expect both $\chi(T)$ and $C_{\rm mag}(T)/T$ to approach
zero for $T \to 0$ at $B\geq 3$ T. From panels (a)-(c) of
Fig.~\ref{fig1}, it is clearly seen that this is not the case,
since for $B\geq 3$ T neither $\chi$ nor $C_{\rm mag}/T$
approach zero as $T\to0$. Moreover, the normalized $C_{\rm
mag}/T$ follows the uniform scaling behavior displayed in
Fig.~\ref{fig1}(c), confirming the absence of a gap.  It is also
seen that the recent measurements of $C_{\rm mag}$
\cite{herb2,herb,t_han:2012,t_han:2014} are compatible with
those obtained on powder samples. These observations support the
conclusions that (i) the properties of $\rm ZnCu_3(OH)_6Cl_2$
under study are determined by a stable SCQSL, and (ii) an
appreciable gap in the spectra of spinon excitations is absent
even under the application of very high magnetic fields of 18 T.
The latter conclusion is in accord with recent experimental
findings that the low-temperature plateau in local
susceptibility identifies the spin-liquid ground state as being
gapless \cite{zorko}.

The same conclusions can be drawn from the neutron-scattering
measurements of the dynamic spin susceptibility $\chi({\bf
q},\omega,T) =\chi{'}({\bf q},\omega,T)+i\chi{''}({\bf
q},\omega,T)$ as a function of momentum $q$, frequency $\omega$,
and temperature $T$. Indeed, these results play a crucial role
in identifying the properties of the quasiparticle excitations
involved. At low temperatures, such measurements reveal that
relevant quasiparticles -- of a new type insulator -- are
represented by spinons, form a continuum, and populate an
approximately flat band crossing the Fermi level
\cite{Han:2012}.  In such a situation it is expected that the
dimensionless normalized susceptibility
$(T^{2/3}\chi'')_{N}=T^{2/3}\chi'' /(T^{2/3}\chi'')_{\rm max}$
exhibits scaling as a function of the dimensionless energy
variable $E_N = E/E_{\rm max}$ \cite{shaginyan:2012:A,book}.
Specifically, the equation describing the normalized
susceptibility $(T^{2/3}\chi'')_{N}$ reads
\cite{shaginyan:2012:A,book}
\begin{equation}\label{SCHIN}
(T^{2/3}\chi'')_N\simeq\frac{b_1E_N}{1+b_2E_N^2},
\end{equation}
where $b_1$ and $b_2$ are fitting parameters adjusted such that
the function $(T^{2/3}\chi'')_{N}$ reaches its maximum value
unity at $E_N=1$ \cite{book,shaginyan:2012:A}. Panel {\bf A} of
Fig.~\ref{fig2} {displays}  $(T^{2/3}\chi'')_{N}$ values
extracted from measurements of the inelastic neutron-scattering
spectrum on the heavy-fermion (HF) metal $\rm
Ce_{0.925}La_{0.075}Ru_2Si_2$ \cite{knafo:2004}. The scaled data
for this quantity, obtained from measurements on two quite
different strongly correlated systems, $\rm ZnCu_3(OH)_6Cl_2$
\cite{herb3} and the deuteronium jarosite $\rm
(D_3O)Fe_3(SO_4)_2(OD)_6$ \cite{faak:2008}, are displayed in
panels {\bf B} and {\bf C} {respectively}. It is seen that the
theoretical results from Ref.~\cite{shaginyan:2012:A} (solid
curves) are in good agreement with the experimental data
collected on all three compounds over almost three orders of {
magnitude of} the scaled variable $E_N$ and hence
$(T^{2/3}\chi'')_{N}$ does exhibit the anticipated scaling
behavior for these systems. From {this observation}  we infer
that the spin excitations in both $\rm ZnCu_3(OH)_6Cl_2$ and
$\rm (D_3O)Fe_3(SO_4)_2(OD)_6$ demonstrate the same itinerate
behavior as the electronic excitations of the HF metal $\rm
Ce_{0.925}La_{0.075}Ru_2Si_2$ and therefore form a continuum.
This detection of a continuum is of great importance since it
clearly signals the presence of a SCQSL in Herbertsmithite
\cite{shaginyan:2012:A,shaginyan:2011:C,book}. It is obvious
from Fig.~\ref{fig2} that the calculations based on this premise
are in good agreement with the experimental data, affirming the
identification of SCQSL as the agent of the low-temperature
behavior of $\rm ZnCu_3(OH)_6Cl_2$ and $\rm
(D_3O)Fe_3(SO_4)_2(OD)_6$. We can only conclude that the spin
gap in the kagome layers is an artificial construction at
variance with known properties of $\rm ZnCu_{3}(OH)_6Cl_2$. In
short, the demonstrable conflicts with experimental data we have
identified negate the existence of a spin gap in the SCQL of
Herbertsmithite. On the other hand, the existence of a gap in
the kagome layers is not in itself of vital importance, for it
does not govern the thermodynamic and transport properties of
$\rm ZnCu_3(OH)_6Cl_2$. Rather, these properties are determined
by the underlying SCQSL. This assertion can be tested by
measurements of the heat transport in magnetic fields, as has
been done successfully in the case of the organic insulators
\cite{yamashita:2010,yamashita:2012,shaginyan:2013:D}.
Measurements of thermal transport are particularly salient in
that they probe the low-lying elementary excitations of SCQSL in
$\rm ZnCu_3(OH)_6Cl_2$ and potentially reveal itinerant spin
excitations that are mainly responsible for the heat transport.
Surely, the overall heat transport is contaminated by the phonon
contribution; however, this contribution is hardly affected by
the magnetic field $B$. In essence, we expect that measurement
of the $B$-dependence of thermal transport will be an important
step toward resolving the nature of the SCQSL in $\rm
ZnCu_3(OH)_6Cl_2$ \cite{shaginyan:2011:C,shaginyan:2013:D,book}.

\begin{figure} [! ht]
\begin{center}
\includegraphics [width=0.38\textwidth]{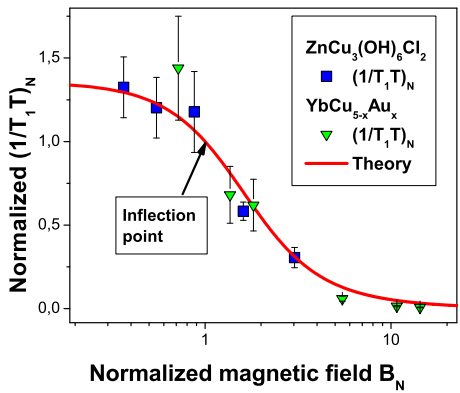}
\end{center}
\caption{ Normalized spin-lattice relaxation rate $(1/T_1T)_N$
at fixed temperature as a function of magnetic field. Data for
$(1/T_1T)_N$ extracted from measurements on $\rm
ZnCu_3(OH)_6Cl_2$ are shown by solid squares \cite{imai} and
those extracted from measurements on $\rm YbCu_{5-x}Au_{x}$ at
$x=0.4$, by the solid triangles \cite{carr}. The inflection
point at which the normalization is taken is indicated by the
arrow. The calculated result is depicted by the solid curve
tracing the scaling behavior of $W_r\propto(M^*)^2$ (see
Eq.~\eqref{WT}).}\label{T12}
\end{figure}

The SCQSL in Herbertsmithite behaves like the {electron liquid}
in HF metals -- provided the charge of an electron is set to
zero. As a result, the thermal resistivity $w=L_T/\kappa$ of the
SCQSL is given by \cite{shaginyan:2011:C,shaginyan:2013:D,book}
\begin{equation}\label{wr}
w-w_0=W_rT^2\propto(M^*)^2T^2,
\end{equation}
where $W_r{T^{2}}$ represents the contribution of spinon-spinon
scattering to thermal transport, being analogous to the
contribution $AT^2$ to charge transport from electron-electron
scattering. (Here $L_T$ is the Lorenz number, $\kappa$ the
thermal conductivity, and $w_0$ the residual resistivity.) Based
on this reasoning it follows that, under application of magnetic
fields at fixed temperature, the coefficient $W_r$ behaves like
the spin-lattice relaxation rate shown in Fig.~\ref{T12}, i.e.,
{ $W_r\propto 1/(T_1T)$}, while in the LFL region at fixed
magnetic field the thermal conductivity is a linear function of
temperature, $\kappa\propto T$
\cite{shaginyan:2011:C,shaginyan:2013:D,book}.

Finally, we consider the effect of a magnetic field $B$ on the
spin-lattice relaxation rate {$1/(T_1T)$}. From Fig.~\ref{T12},
which shows the normalized spin-lattice relaxation rate {
$1/(T_1T)_N$} at fixed temperature versus magnetic field $B$, it
is seen that increasing $B$ progressively reduces { $1/(T_1T)$},
and that as a function of $B$, there is an inflection point at
some $B=B_{\rm inf}$, marked by the arrow. To clarify the
scaling behavior in this case, we normalize { $1/(T_1T)$} by its
value at the inflection point, while the magnetic field is
normalized by $B_{\rm inf}$.  Taking into account the relation
$1/(T_1T)_N\propto(M^*)^2$, we expect that a strongly correlated
Fermi system located near its quantum critical point will
exhibit the similar behavior of { $1/(T_1T)_N$}
\cite{pr,shaginyan:2011:C,shaginyan:2013:D,book}. Significantly,
Fig.~\ref{T12} shows that Herbertsmithite $\rm ZnCu_3(OH)_6Cl_2$
\cite{imai} and the HF metal $\rm YbCu_{5-x}Au_{x}$ \cite{carr}
do show the same behavior for the normalized spin-lattice
relaxation rate. As seen from Fig.~\ref{T12}, for $B\leq B_{\rm
inf}$ (or $B_N\leq1$) the normalized relaxation rate {
$1/(T_1T)_N$} depends weakly on the magnetic field, while it
diminishes at the higher fields
\cite{pr,shaginyan:2011:C,shaginyan:2013:D,book} according to
\begin{equation}\label{WT}
W_r\propto{ 1/(T_1T)_N}\propto(M^*)^2\propto B^{-4/3}.
\end{equation}
Thus, we conclude that { the} application of a magnetic field
$B$ leads to a crossover from  NFL to LFL behavior and to a
significant reduction in both the relaxation rate and the
thermal resistivity.

In summary, we have demonstrated that both the impurity model of
Herbertsmithite and the existence of a spin gap are problematic,
as they contradict established properties of Herbertsmithite and
are not supported by considerations of the thermodynamic and
relaxation properties in magnetic fields. We conclude by
recommending that measurements of heat transport in magnetic
fields be carried out to clarify the quantum spin-liquid physics
of Herbertsmithite $\rm ZnCu_{3}(OH)_6Cl_2$.

\end{document}